# The Coulomb-Higgs transition of the three-parameter U(1)-Higgs model

A. Cruz[a], D. Iñiguez[a], A. Tarancón[a] , L. A. Fernández[b] and A. Muñoz Sudupe [b*]

[a]Departamento de Física Teórica, Universidad de Zaragoza, 50009 Zaragoza, Spain

[b]Departamento de Física Teórica, Universidad Complutense de Madrid, 28040 Madrid, Spain

We find a first order Coulomb–Higgs phase transition at moderately large values of the coupling $\lambda$, and no evidence for a change of order at any finite value of it.

## 1. INTRODUCTION

The Higgs mechanism is an essential part of the present day formulation of the Standard Model. The U(1) gauge model coupled to scalars is a simplification of the gauge–Higgs sector of the SM, which yet keeps the unsolved problem of defining non perturbatively a non–asymptotically free field theory, and which can be addressed with rather simple technical means.

## 2. THE COULOMB–HIGGS PHASE TRANSITION IN THE $\beta$ $\kappa$ MODEL

The three parameter U(1)–Higgs model is described by the action

$$S = -\beta \sum_{\mathbf{r},\mu<\nu} \Re U_{\mathbf{r},\mu\nu} - \kappa \sum_{\mathbf{r},\mu} \Re \bar{\Phi}_{\mathbf{r}} U_{\mathbf{r},\mu} \Phi_{\mathbf{r}+\mu} + \lambda \sum_{\mathbf{r}} (|\Phi_{\mathbf{r}}|^2 - 1)^2 + 4\kappa \sum_{\mathbf{r}} |\Phi_{\mathbf{r}}|^2 \quad (1)$$

In the $\lambda \to \infty$ limit, $|\Phi| \to 1$ and the action simplifies to

$$S = -\beta \sum_{\mathbf{r},\mu<\nu} \Re U_{\mathbf{r},\mu\nu} - \kappa \sum_{\mathbf{r},\mu} \Re \bar{\Phi}_{\mathbf{r}} U_{\mathbf{r},\mu} \Phi_{\mathbf{r}+\mu} \quad (2)$$

The phase diagram of that restricted version [1] has been discussed by us in Latt93 [2], and the difficulty to identify the order of the Coulomb–Higgs phase transition was made apparent. The $\beta \to \infty$ end of that line is the X–Y model phase transition, a well established second order one, which has induced researchers to think the line to be a prolongation of its end point, and as such, second order [4]. Yet, the relationship of the model to the $q > 1$ version and to the $Z_N$ model reported in [2] and [3] points towards a possible first order transition.

## 3. THE THREE PARAMETER MODEL

The difficulty to study directly the $\lambda \to \infty$ limit has lead us to study the problem at finite, variable $\lambda$. This had been done earlier [5], and we intend to improve on the statistics. In order to achieve that goal, we have fixed $\beta = 1.15$, which is on the Coulomb–Higgs side of the first order Confining–Coulomb phase transition. We have worked at values of $\lambda$ $0.003, 0.005, 0.01, 0.03, 0.1, 0.3$ on lattices $6^4, 8^4, 12^4$ and $16^4$, with statistics ranging between $10^5$ and $10^6$ Montecarlo iterations per measurement, on workstations, on parallel machines at the Edinburgh Parallel Computing Centre and on our custom, 64 T800 processor computer RTN [6]. We have used an over-relaxed Metropolis update method, with a multicanonical procedure to accelerate flip-flop rate when the latter is too low. The jack-knife method has been used in the error estimation.

## 4. THE ENERGY HISTOGRAMS

Fig. 1 shows typical histograms of the link energy $\langle \bar{\Phi} U \Phi \rangle$, whose most striking features are:

- Two-peak structure.
- Assymetry of the peaks, with a narrow low-energy peak and a broad high-energy one.

*Partial support is acknowledged from CICYT contracts AEN93 0604, AEN93 0776 and AEN94 0218 and from EC contract CHRX-CT92-0051. D.I. acknowledges MEC, CAI and Edinburgh University.

- Strong size dependence, consisting on a narrowing of the gap and width of the peaks as the lattice size $L$ increases, effected by a desplacement towards lower energy of both peaks, especially the broad one.

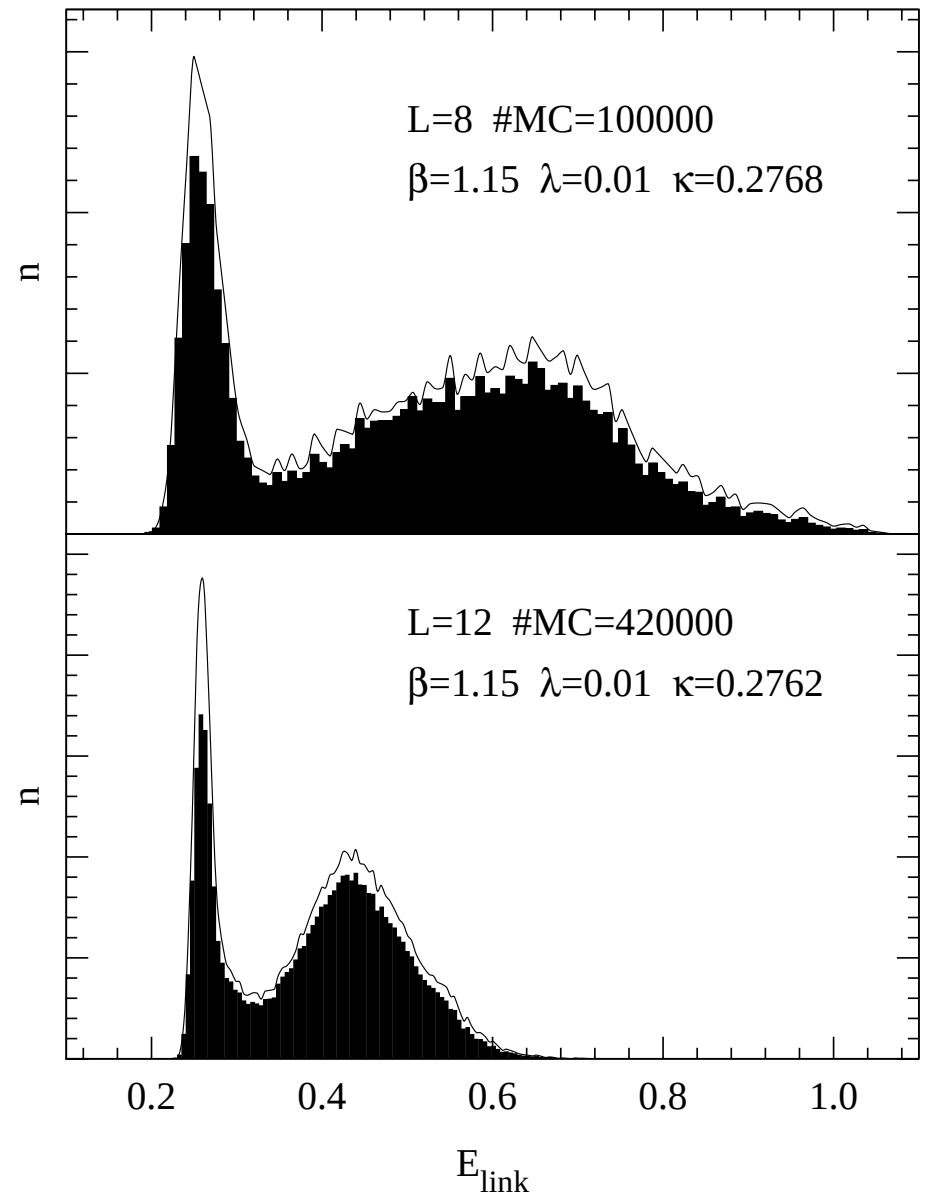

Figure 1. Histograms of the link energy in an $8^4$ lattice and a $12^4$ lattice at $\lambda = 0.01$

## 5. MEASUREMENTS OF THE LATENT HEAT

In order to quantify the general features listed in the previous section, we have measured the latent heat at the apparent critical point at a given $L$, defined as the value $\kappa_c$ for which the link-energy spread (specific heat) is maximum. The definition is global, and so, little dependent on local details, and has a clear thermodynamic limit.

The latent heat has been defined as the difference between the positions of the maxima for a fit to each peak separately to a cubic spline, after the histogram has been shifted to the apparent critical point by the spectral density method [7]. A cubic spline has been preferred to other functional shapes because of its ability to reproduce a maximum and accomodate the mixed states, whose influence on the histogram is otherwise difficult to account for.

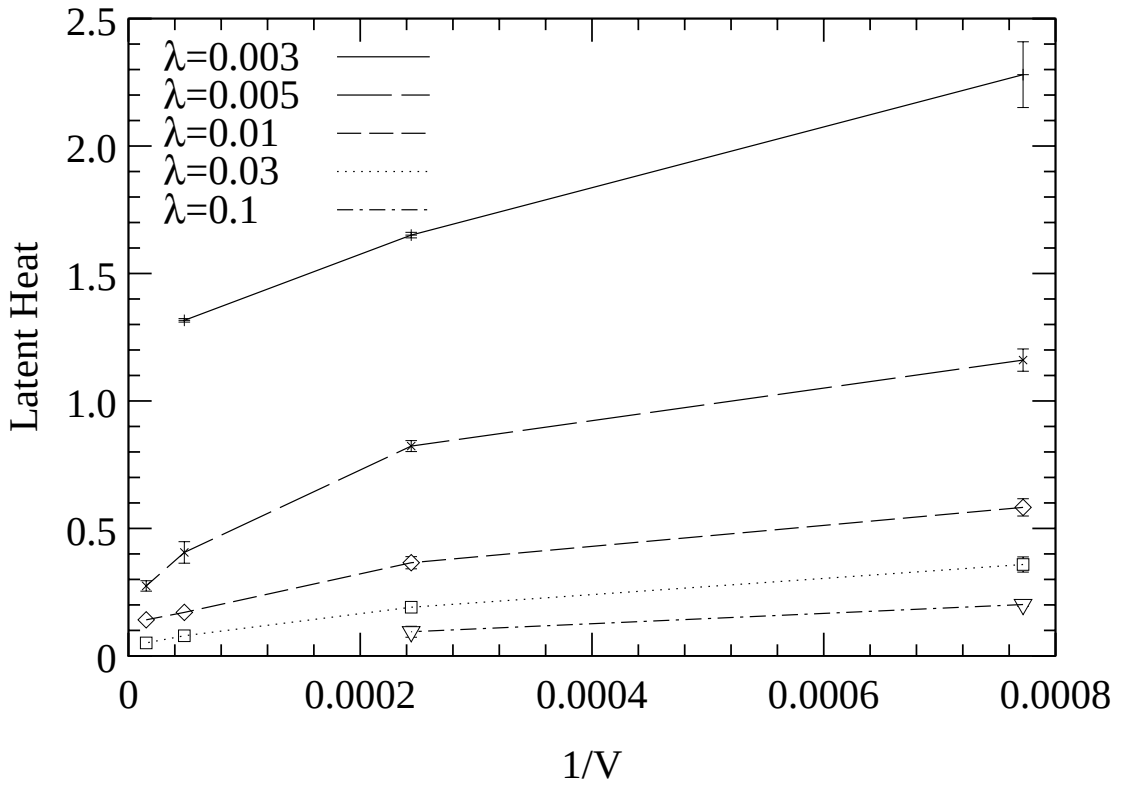

Figure 2. Latent heat as a function of $1/V$ at different values of $\lambda$

Fig. 2 shows, for $\lambda \leq 0.03$, evidence for a non-zero-point extrapolation of the latent heat in the $V \to \infty$ limit, suggesting the first order of the phase transition for smaller values of $\lambda$. For $\lambda > 0.03$ the figure is not conclusive, so that on the basis of the behaviour of the latent heat alone, the ocurrence of a tricritical point at $\lambda > 0.03$ cannot be descarded.

## 6. THE MAGNETIC SUSCEPTIBILITY

In analogy with magnetic systems, we define the magnetic susceptibility as $\chi = \sigma^2 V$, $\sigma^2$ being the variance of each peak in the histogram. $\sigma$ is measured by fitting the log of the energy around each peak to a parabola, and the dependence of $\sqrt{\chi}$ on $1/V$ is shown in Fig. 3.

The figure shows a rather flat behaviour of the susceptibility as $V \to \infty$ at all values of $\lambda$, with no trace of divergence as $\lambda$ increases. This behaviour seems to exclude the existence of a second order phase transition in any limit, pointing to the possibility of the first order character being kept in the thermodynamic limit as $\lambda \to \infty$.

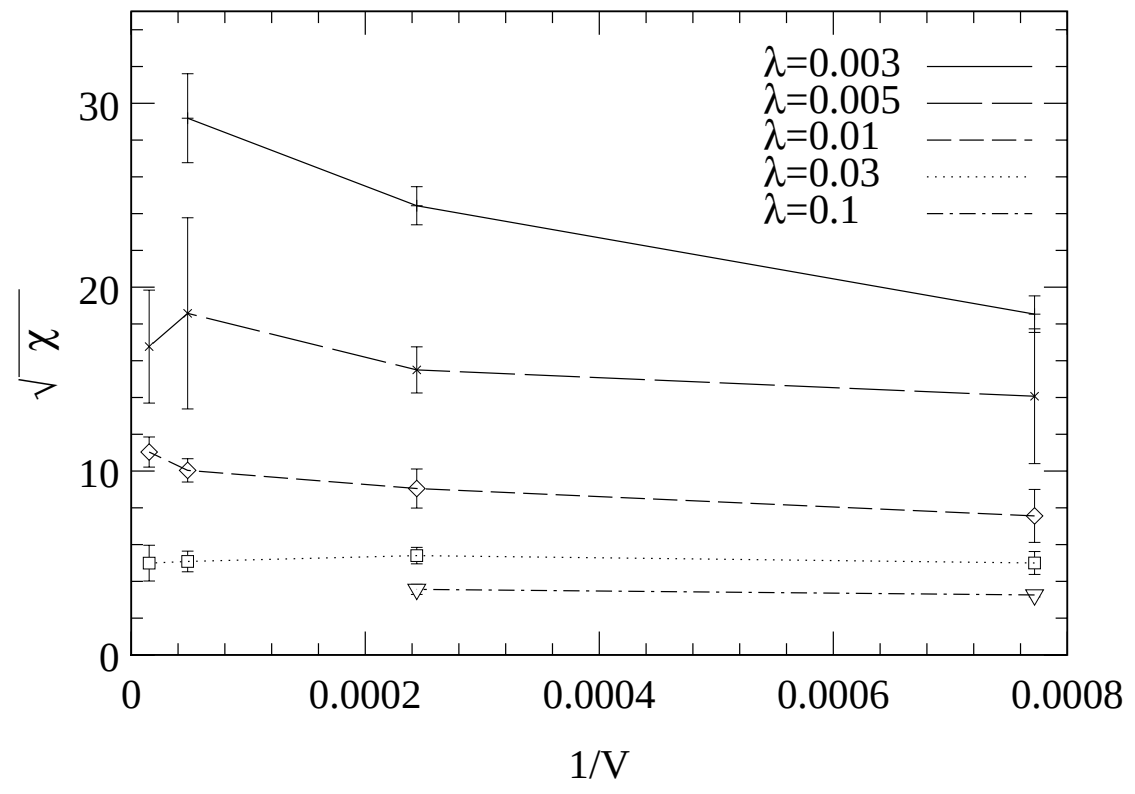

Figure 3. Square root of the susceptibility of the higher state versus $1/V$ at different values of $\lambda$

## 7. FINITE SIZE EFFECTS

The finite size effects observed can be accounted for almost completly by the shift of the apparent critical point produced by the change of lattice size. Fig. 4 shows the position of the high energy peak when the critical point, $\kappa_c$, is determined as the value for which the specific heat is maximum, which are the values used all through the paper, or as the value at which the two peaks have equal hight, another popular choice for the definition of the apparent critical point. Also plotted is the position of the peak when, at different lattice sizes $L$, the value of $\kappa_c$ measured at $L = 16$ is used. As can be seen, the position of the peak does not change practically.

## 8. CONCLUSIONS

The analysis of the latent heat shows that the Coulomb-Higgs phase transition is first order for all the finite values of $\lambda$ studied. The added evidence of the magnetic susceptibility not growing with the lattice size and with $\lambda$ seems to exclude the existence of a finite $\lambda_c$ beyond which the transition becomes second order and points towards the possibility of it being first order even at $\lambda = \infty$ (fixed modulus case).

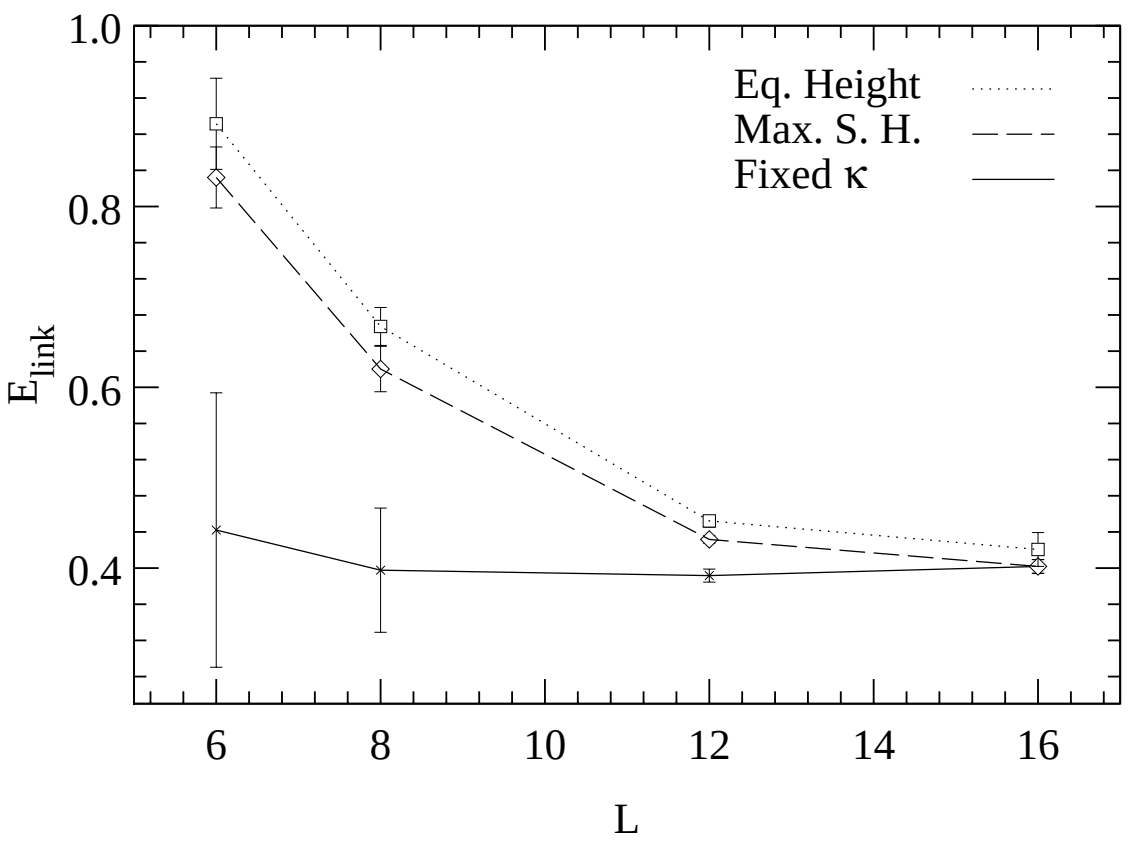

Figure 4. Positions of the high energy peak as three different definitions of $\kappa_c$ (maximum specific heat, equal hight and $\kappa_c$ at $L = 16$ ) are used. $\lambda$ is 0.01